\documentclass[10pt,preprint]{aastex}

\RequirePackage{graphicx}
\RequirePackage{color}
\RequirePackage{code}
\input{astro.sty}

\usepackage[T1]{fontenc}

\def\plotext{pdf}
\def\picdir{pics}

\shorttitle{Charge Diffusion Variations in PS1}
\shortauthors{E.A. Magnier et al}
\begin{document}
\title{Charge Diffusion Variations in Pan-STARRS\,1 CCDs}

\def\IfA{1}
\def\CfA{2}
\def\LBL{3}
\def\Hubble{4}
\def\DUH{5}

\author{
Eugene A. Magnier,\altaffilmark{\IfA}
J.~L. Tonry, \altaffilmark{\IfA}
D. Finkbeiner,\altaffilmark{\CfA}
E. Schlafly,\altaffilmark{\LBL,\Hubble}
W.~S. Burgett,\altaffilmark{\IfA}
K.~C. Chambers,\altaffilmark{\IfA} 
H.~A. Flewelling,\altaffilmark{\IfA}
K. W. Hodapp,\altaffilmark{\IfA}
N. Kaiser,\altaffilmark{\IfA}
R.-P. Kudritzki,\altaffilmark{\IfA}
N. Metcalfe,\altaffilmark{\DUH}
R. J. Wainscoat,\altaffilmark{\IfA} and 
C. Z. Waters,\altaffilmark{\IfA}
} 

\altaffiltext{\IfA}{Institute for Astronomy, University of Hawaii, 2680 Woodlawn Drive, Honolulu HI 96822}
\altaffiltext{\CfA}{Harvard-Smithsonian Center for Astrophysics, 60 Garden Street, Cambridge, MA 02138}
\altaffiltext{\DUH}{Department of Physics, Durham University, South Road, Durham DH1 3LE, UK}
\altaffiltext{\LBL}{Lawrence Berkeley National Laboratory, One Cyclotron Road, Berkeley, CA 94720, USA}
\altaffiltext{\Hubble}{Hubble Fellow}
\begin{abstract}

Thick back-illuminated deep-depletion CCDs have superior quantum
efficiency over previous generations of thinned and traditional thick
CCDs.  As a result, they are being used for wide-field imaging cameras
in several majorxs projects.  We use observations from the Pan-STARRS
$3\pi$ survey to characterize the behavior of the deep-depletion
devices used in the Pan-STARRS\,1 Gigapixel Camera.  We have
identified systematic spatial variations in the photometric
measurements and stellar profiles which are similar in pattern to the
so-called ``tree rings'' identified in devices used by other
wide-field cameras (e.g., DECam and Hypersuprime Camera).  The
tree-ring features identified in these other cameras result from
lateral electric fields which displace the electrons as they are
transported in the silicon to the pixel location.  In contrast, we
show that the photometric and morphological modifications observed in
the GPC1 detectors are caused by variations in the vertical charge
transportation rate and resulting charge diffusion variations.
\end{abstract}


\section{INTRODUCTION}\label{sec:intro}

CCD detectors have evolved greatly since they were first introduced
for astronomical imaging in the mid 1970s.  In addition to the
well-known increases in the size of CCDs over the past 4 decades, CCD
architecture has gone through three major evolutionary stages.  

The first generation of CCDs used a silicon substrate a few hundred
microns thick on top of which gate structures were deposited to define
the pixels.  A positive voltage applied to the gate layers would
create a shallow region (\approx 10 microns thick) in which the holes
were depleted.  This ``depletion region'' acted as a potential well to
trap electrons, specifically those generated by absorbed photons.  The
thick silicon substrate required illumination from the ``front'' side
containing the thin gate structures to allow the photons to reach the
depletion region and be detected.  These early CCDs had modest quantum
efficiency as photons were easily absorbed by the several-micron-thick
gate structures.  For an excellent review of the history of CCD
development, see \cite{1992ASPC...23....1J}.

Thinned, backside-illuminated CCDs such as the TI 3PCCD
\citep{1981SPIE..290....6B} were developed to address the quantum
efficiency limitations of the first generation thick CCDs.  The
silicon substrate was removed using a chemical process, leaving a
delicate device only \approx 10 - 20\micron\ thick, exposing the
depletion region on the backside.  Photons entering the backside of
the device are not blocked by the gate structures and are thus more
easily absorbed and detected.  Thinned backside-illuminated CCDs have
high quantum efficiency to blue photons.  However, as the wavelength
increases beyond \approx 800 nm, the silicon becomes more transparent
to the photons with a corresponding drop in quantum efficiency for
red photons.  In addition, thin-film interference between the entering
photons and those reflecting off the front side of the CCD result in
``fringe'' patterns for redder photons.

Early generations of CCDs were made of low-resistivity (\approx 10 -
50 $\Omega$-cm) silicon.  Following experiments beginning in the early
1990s \citep{Holland.1996}, CCDs made from thick, high-resistivity ($
> 10 k\Omega$-cm) silicon were developed for astronomical instruments
in the early 2000s \citep{Holland.2003}.  The high-resistivity of the
silicon allows for depletion regions of hundreds of microns in depth,
compared to \approx 10\micron\ for the low-resistivity silicon.  This
modification allows for a back-illuminated CCD with a relatively thick
silicon subtrate of 75 - 300\micron.  Blue photons impinging on the
back of the device are absorbed near the back surface of the device
and are caried through the depletion region to the gates on the front
side.  The thick silicon allows red photons to have a greater chance
to be absorbed, increasing quantum efficiency in the red.  Because
these thick, deep-depletion devices have near-unity quantum efficiency
across a very wide spectral range, they have become the design of
choice for many modern, large-scale CCD cameras (e.g., Pan-STARRS
GPC1, \citealt{2009amos.confE..40T}; Subaru Hypersuprime Camera,
\citealt{2010SPIE.7735E..3FK}; Dark Energy Survey Camera,
\citealt{2015AJ....150..150F}).

While these deep-depletion CCDs seem to be ideal, they do have
features which can cause challenges for precise measurements.  For
example, as a result of the ``Brighter-Fatter Effect''
\citep{2014JInst...9C3048A,2015JInst..10C5032G}, the profile of bright
stars are measured to be wider than the profiles of faint stars.  The
accepted interpretation is that the electric fields produced by the
electrons accumulated from a star repel successive incoming electrons,
with the repulsion increasing the more electrons have accumulated.

The effects of lateral electric fields are likewise identified as the
cause of the so-called ``tree rings'' observed in the flat-field,
astrometry, and photometry response of thick deep-depletion detectors
\citep{2014PASP..126..750P}.  These tree-ring patterns have been noted
in the flat-field response of deep depletion devices since their early
testing \citep[see, e.g., Figure 2 in][]{2010SPIE.7735E..1RE} and were
initially considered to be a sensitivity response which could be
removed with a flat-field.  As discussed in detail by
\cite{2014PASP..126..750P}, these tree rings are more correctly
interpretted as variations in the effective pixel area due to
migration of the electrons pushed by lateral electric fields induced
by small changes in the doping used to set the resistivity of the
silicon.  The changes in the effective area result in changes to the
apparent flat-field response as well as the astrometric response of
the detector.  More subtly, the changes in the flat-field response,
since they do not reflect actual variations in sensitivity, can lead
to systematic photometry errors for astronomical sources if flat-field
images are used in the standard fashion.

In this paper, we examine the behavior of an apparently-similar kind
of tree-ring pattern observed in the Pan-STARRS\,1 Gigapixel Camera 1
CCDs.  Although we also observe the changes in effective pixel area
caused by lateral electric fields as described by
\cite{2014PASP..126..750P}, we show below a second effect which is
more important in these devices in driving systematic photometry
errors.  We find that variations in charge diffusion, also resulting
from changes in the silicon doping structures, affect both the
observed stellar profiles as well as the photometry measured with
profile fitting techniques.  In Section~\ref{sec:PS1}, we discuss the
Pan-STARRS\,1 telescope, camera, and survey data used in this analysis.
In Section~\ref{sec:tree.rings}, we present the tree-ring patterns as
observed in several different types of measurements: flat-field
response, systematic photometric residuals, systematic astrometric
residuals, and stellar profile shape variations.  In
Section~\ref{sec:discussion}, we discuss the interpretation of
patterns we observe and present a simple model to explain the observed
behavior.  We conclude with a discussion of the implications of this
effect on astronomical measurements from deep depletion instruments

\section{Pan-STARRS1}
\label{sec:PS1}

The 1.8m Pan-STARRS\,1 telescope (PS1), located on the summit of
Haleakala on the Hawaiian island of Maui, has been surveying the sky
regularly since May 2010 \citep{chambers2017}.  From May 2010 through
March 2014, PS1 was run under the aegis of the Pan-STARRS\,1 Science
Consortium (PS1SC) to perform a set of wide-field science surveys;
since March 2014, operations have been supported primarily by NASA's
Near Earth Object Observation program
\citep[see][]{2015IAUGA..2251124W}.  Under the PS1SC, the largest
survey, both in terms of area of the sky covered ($3\pi$ steradians)
and fraction of observing time (56\%), was the \TPS\ in which the
entire sky north of Declination $-30$\degrees\ was imaged \approx 80
times over 4 years.  These observations were distributed over five
filters, \grizy, and have been astrometrically and photometrically
calibrated to good precision \citep{magnier2017.calibration}.

The wide-field PS1 telescope optics \citep{2004SPIE.5489..667H} image
a 3.3 degree field of view on a 1.4 gigapixel camera
\citep[GPC1;][]{2009amos.confE..40T}, with low distortion and generally
good image quality.  The median seeing for the \TPS\ data vary
somewhat by filter: (\grizy) = (1.31, 1.19, 1.11, 1.07, 1.02)
arcseconds.  Routine observations are conducted remotely from the
Advanced Technology Research Center in Kula, the main facility of the
University of Hawaii's Institute for Astronomy operations on Maui.

GPC1, currently the largest astronomical camera in terms of number of
pixels, consists of a mosaic of 60 edge-abutted $4800\times4800$ pixel
detectors, with 10~$\mu$m pixels subtending 0.258~arcsec. These CCID58
detectors, manufactured by Lincoln Laboratory, are 75\micron-thick
back-illuminated CCDs \citep{2006amos.confE..47T,2008SPIE.7021E..05T}.
Initial performance assessments are presented in
\cite{2008SPIE.7014E..0DO}. The active, usable pixels cover \approx
80\% of the FOV.

\subsection{Data Processing and Calibration}


Images obtained by PS1 are processed by the Pan-STARRS Image
Processing Pipeline (IPP;
\citealp{2006amos.confE..50M,magnier2017.datasystem}).  All
observations are processed nightly, with results sent to groups within
the science consortium (i.e., PS1SC during the \TPS) performing
short-term science projects (e.g., searching for transient and moving
objects).  In addition, the \TPS\ dataset has been re-processed
several times with improved calibration and analysis techniques.  To
date (2017 September), 3 re-processings starting from raw pixel data
have been performed.  The labels PV0, PV1, PV2, PV3 are used identify
the nightly processing and successive re-processing versions.  PV3 has
been used for the public release of the Pan-STARRS \TPS\ data via the
{\it Barbara A. Mikulski Archive for Space Telescopes} (MAST) at the
Space Telescope Science Institute.\footnote{http//panstarrs.stsci.edu}
The process of the construction of this database and the schema
details are discussed in detail by \cite{flewelling2017}.

The data processing and calibration operations are discussed in detail
in elsewhere
\citep{magnier2017.analysis,magnier2017.calibration,waters2017}.
We re-visit here a number of points that are of significance to this
study.  Images are processed following a fairly standard sequence of
image detrending, source detection, and initial calibration
(astrometric and photometric) of those detected sources.  Additional
standard processing critical to PS1 science operations includes
geometric transformation (`warping') and image combinations (summed
stacks and differences).  For the purposes of this analysis, we are
only considering the sources detected in the individual exposures from
the initial analysis steps.



As discussed in \cite{waters2017}, image detrending includes
flat-field processing with a single epoch flat-field image for each
filter.  The flat-field image used for this analysis has been
generated by median-combining dome flat-field images (after
pre-processing and pixel outlier rejections) and then multiplying by a
photometric flat-field correction image generated by the analysis of a
grid of images of a dense stellar field.  The purpose of this second
step is to correct the basic flat-field image for errors arising from
the non-uniformity of the illumination, from non-pixel uniformity due
to the varying optical distorition across the field, and any other
factors which may make the flat-field image inconsistent with stellar
photometry, e.g., SED, filter band-pass variations, etc
\citep[see][]{waters2017,2004PASP..116..449M,2007ASPC..364..153M}.
This correction was made on a relatively coarse grid across the focal
plane in order to accumulate sufficient statistics from the stars in
the relatively small number of images available at the time.  We have
found that a single flat-field set can be used for all PS1
observations to yield photometric systematic errors at the level of \approx
2\%.  PS1 benefits in this regard from the stability of having a
single instrument which is rarely removed.

Photometry of the PS1 images is performed using a
point-spread-function (PSF) model as well as multiple kinds of
apertures \citep{magnier2017.analysis}.  In this analysis, we refer to
aperture photometry performed using an aperture defined based on the
image quality observed for a given chip.  The aperture diameter is set
to be 3.75 times the FWHM for the image.

To improve the photometric systematic errors beyond the level achieved
with a single (photometrically corrected) flat-field set, the PS1
photometry is re-calibrated within the databasing system based on the
properties of the measured photometry.  The calibration process is
discussed by \cite{2012ApJ...756..158S} and
\cite{2013ApJS..205...20M,magnier2017.calibration}.  As part of this
process, several flat-field corrections have been determined.  For the
PV2 analysis discussed here, a flat-field correction determined during
the ubercal analysis \citep[see][]{2012ApJ...756..158S} consisted of
an $8\times 8$ grid of corrections for each GPC1 chip, corresponding
to a correction for each OTA ``cell'' and filter for each of 4
seasons.  The boundaries of those seasons are tentatively identified
with modifications to the baffle structures or the system optics.  The
critical point here is that the final effective flat-field image for
the PV2 dataset is based on a dome-flat at the highest resolution,
with very low resolution (hundreds of pixels) corrections based on
photometry, resulting in photometric systematic uncertainties in the
range 7 - 12 millimagnitudes, depending on the filter
\citep{2013ApJS..205...20M}.

For all objects, positions are measured from the PSF model for the
brighter sources (using a non-linear fitting process) and from a
simple centroid (1st moment) for the fainter source
\citep{magnier2017.analysis}.  These position measurements are
used in the astrometric analysis.  The astrometric calibration is
discussed by \cite{magnier2017.calibration}; for the PV2
dataset, the typical systematic floor is \approx 15 - 20
milliarcsecond for individual measurements of brighter stars. 

\section{Tree-Ring Patterns}
\label{sec:tree.rings}

\begin{table}
\begin{center}
\caption{Systematic Trends : Standard deviation by filter\label{table:sigmas.by.filter}}
\begin{tabular}{|l|rrrrr|}
\hline
{\bf Filter} & {\bf psf mags} & {\bf ap mags} & {\bf astrom} & {\bf smear} & {\bf flat} \\
             & mmags         & mmags          & mas          & pixels$^2$  & mmags \\
\hline
\gps & 11.8 & 13 & 8.0  & 0.169 &  3.0 \\ 
\rps & 10.9 & 12 & 6.7  & 0.133 &  2.2 \\
\ips &  8.5 & 10 & 6.0  & 0.069 &  1.7 \\
\zps &  8.7 & 12 & 5.5  & 0.052 &  3.2 \\
\yps & 16.5 & 26 & 6.8  & 0.059 & 15.3 \\
\hline
\end{tabular}
\end{center}
\end{table}

For many of the GPC1 OTA CCDs, we observe a spatial pattern in the
photometric residuals for each device which is similar in appearence
to the tree rings described in the Dark Energy Camera (DECam) by
\cite{2014PASP..126..750P}.  This pattern consists of systematic
deviations which are consistent in a set of circular arcs centered on
the corner of the CCD, as shown in Figure~\ref{fig:psfmags.by.filter}.
The details of the analysis used to generate
Figure~\ref{fig:psfmags.by.filter} are given below.  For now, we note
that the GPC1 CCDs are constructed by dividing the circular silicon
wafer into 4 inscribed squares.  Thus the corners of the CCDs lie in
the center of the silicon boule, just as the center of the circular
tree rings described by \cite{2014PASP..126..750P} match the center of
the boule from which they came.  This gives the impression that a
similar mechanism is responsible for the pattern observed in the PS1
photometry and the DECam photometry, namely the diffusive effects of
lateral electric field variations in the detectors.  In the next
section, we will make the case that the patterns observed in the PS1
photometry residuals are {\em not} caused by this mechanism, but are
instead caused by variations in the {\em vertical} electric field (the
field direction perpendicular to the CCD surface).

First, in this section, we will describe how we have measured the
presence or absence of these tree-ring patterns in 5 types of data.
For all of these examples, we use a single GPC1 CCD (XY40) to
illustrate the effects in detail, but a similar set of effects are
seen in many, if not all, of the GPC1 detectors with varying
strengths.  First, we show the residual PSF photometry.  Second, we
show the residual aperture photometry.  Third, we show the astrometric
residual patterns.  Fourth, we show the patterns observed in the
flat-field images.  Finally, we show measurements derived from the
second-moments of the stars.

For all effects discussed below, we are measuring the mean value of
the effect in 10x10 pixel superpixels across the detector.  The
resulting images are all constructed so that a given superpixel
represents the same range of true GPC1 XY40 pixels regardless of the
type of measurement.  To generate the photometry, astrometry, or
second-moment plots, measurements were extracted from the PV0 DVO
database \citep{magnier2017.calibration} for observations covering
the region ($\alpha$,$\delta$) = (90\degree\ -- 150\degree,
-25\degree\ -- 10\degree).  This region of the sky provides a fairly
high density of stars, but avoids the Galactic Plane where confusion
may potentially contaminate the measurement.  We limit the analysis to
good measurements (\ippmisc{PSF_QF} $>$ 0.85, see
\citealt{magnier2017.analysis}) of likely stars ($|m_{psf} -
m_{aper}| < 0.2$).  Only measurements with instrumental magnitude $<
-8.0$ ($-2.5\log \mbox{cts sec}^{-1} < -8.0$) are included to ensure
reasonable signal-to-noise per measurement.  We require at least 2
measurements in a given filter and at least 5 measurements total for
any star included in the analysis.

\subsection{Photometric Residuals}

\def\figwidth{5.2in}
\def\jumpleft{-2.6in}
\def\capwidth{2.4in}
\begin{figure*}[htbp]
\begin{center}
\parbox[b]{\figwidth}{\includegraphics[width=\figwidth]{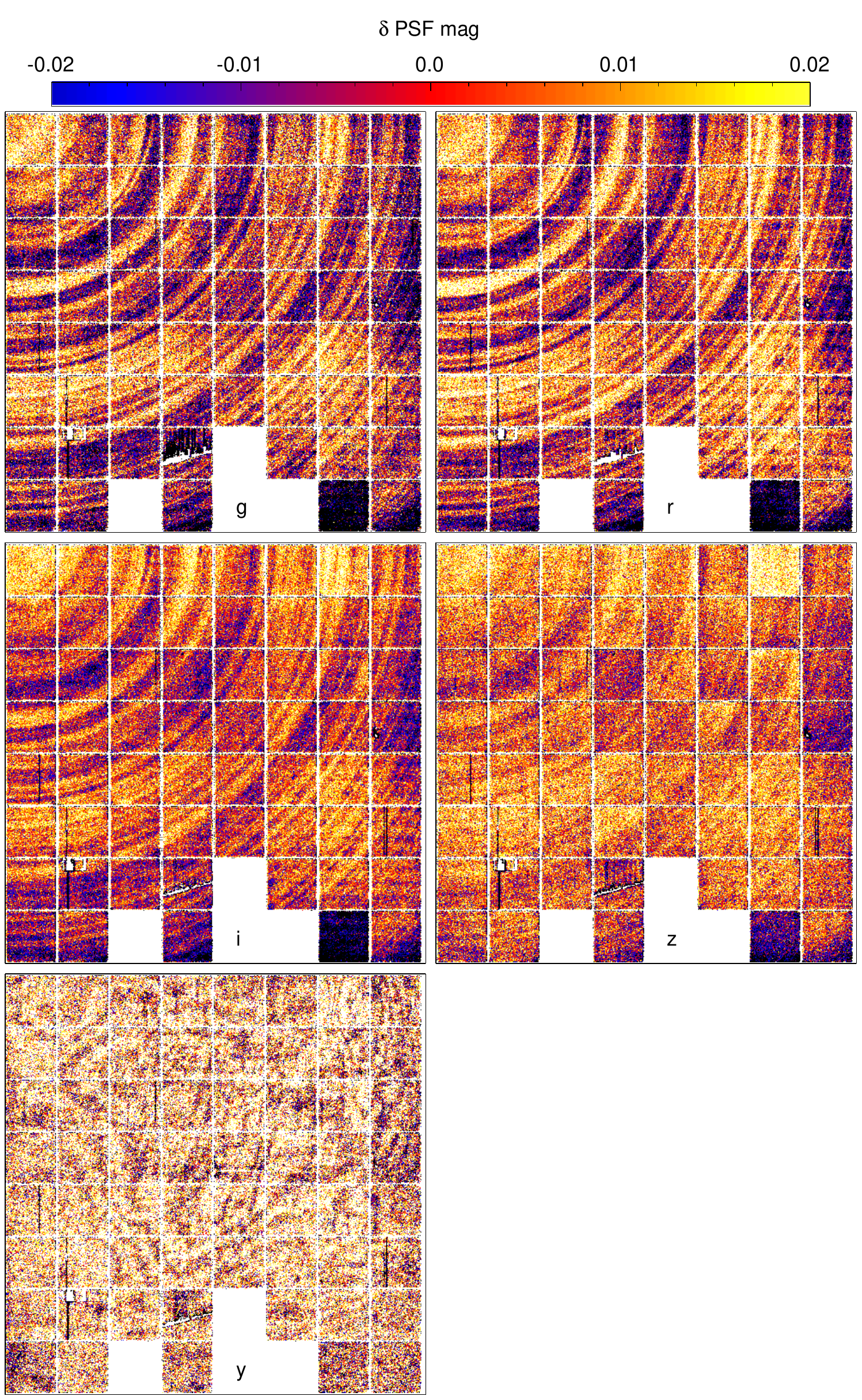}}
\hspace{\jumpleft}
\parbox[b]{\capwidth}{
\caption{PSF Magnitude residuals by filter (\grizy).  White boxes are
  GPC1 cells which have been masked due to poor response.  Superpixels
  representing regions of $10\times10$ pixels are used to determine
  the median deviation for measurements at the given chip pixel
  location compared with the average photometry for the given
  object.} \label{fig:psfmags.by.filter}}
\end{center}
\end{figure*}

\begin{figure*}[htbp]
\begin{center}
\parbox[b]{\figwidth}{\includegraphics[width=\figwidth]{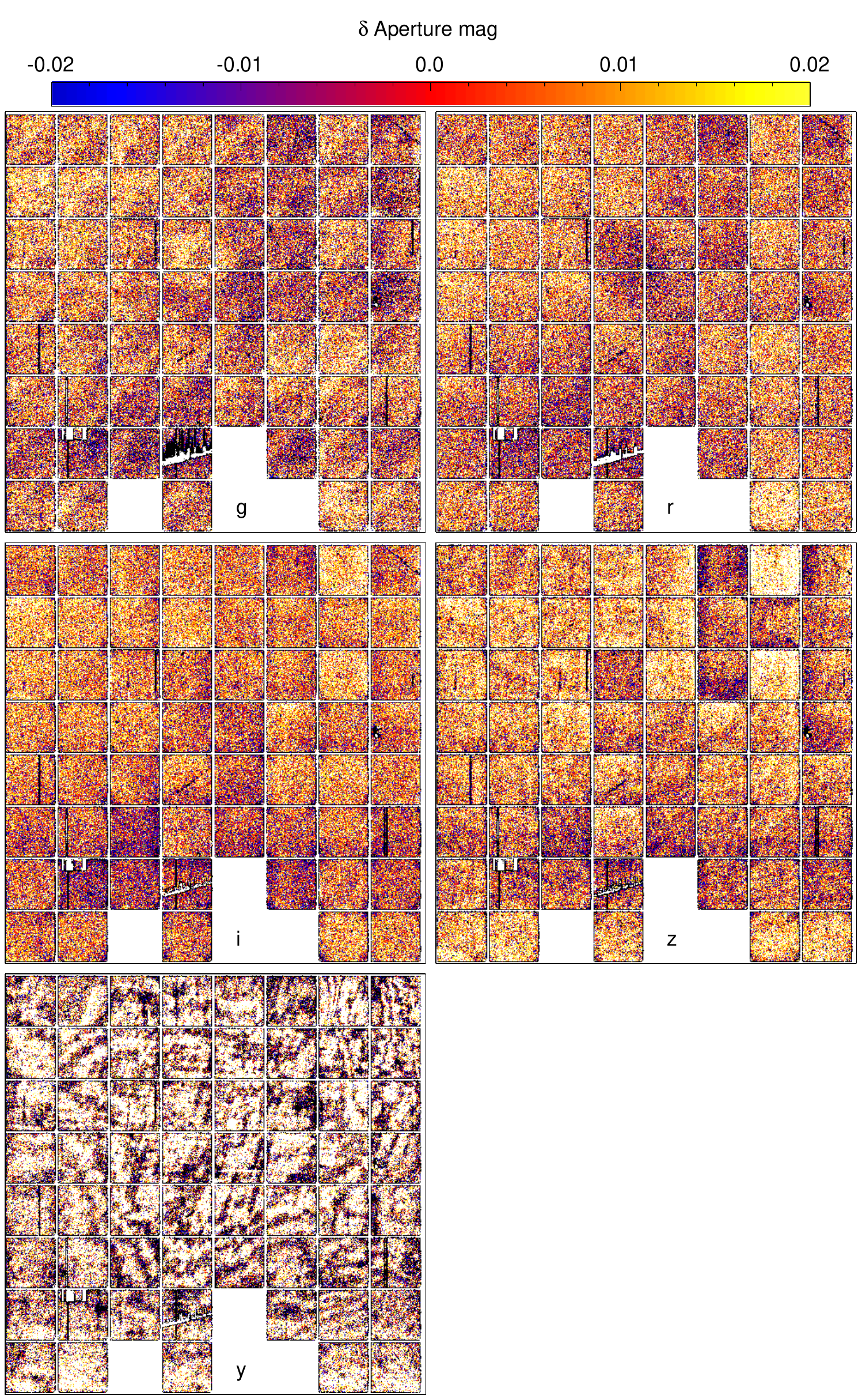}}
\hspace{\jumpleft}
\parbox[b]{\capwidth}{
\caption{Aperture Magnitude residuals by filter (\grizy).  White boxes
  are GPC1 cells which have been masked due to poor response.
  Superpixels representing regions of $10\times10$ pixels are used to
  determine the median deviation for measurements at the given chip
  pixel location compared with the average photometry for the given
  object.  } \label{fig:apmags.by.filter}}
\end{center}
\end{figure*}

Figure~\ref{fig:psfmags.by.filter} shows the 2D patterns of PSF
photometry residuals.  In this case, we select PSF magnitude
measurements for detections of stars which fall in the given
superpixel.  We subtract each measurement from the average magnitude
for the object in the selected filter ($\delta m_{psf} =
\overline{m}_{psf} - m_{psf}$) to determine the residual magnitude,
excluding as an outlier any measurement with $|\delta m_{psf}| > 0.5$.
For a given superpixel, we measure the median of the $\delta m_{psf}$
distribution.  The figure shows $\delta m_{psf}$ for each filter
(\grizy).  The dynamic range of the color scale is from -20 to +20
millimagnitudes for all 5 plots.

The tree-ring pattern is clearly visible for the four blue filters,
but finging dominates the pattern for \yps.  Small offsets of
individual cells are also apparent for \zps.  While the patterns are
clear across the image, the signal-to-noise of the structures per
pixel is not very strong in these images.  The per-pixel standard
deviations of these plots are listed in
Table~\ref{table:sigmas.by.filter}.  The per-pixel standard deviation
is comparable to the amplitude of the correlated structures, so we
need to integrate along the radial structures to make stronger
statements about these patterns.

Figure~\ref{fig:apmags.by.filter} shows the equivalent measurement for
aperture photometry instead of PSF photometry.  The finging
pattern again dominates the plot for \yps, but the tree rings are not
seen in any of the filters.  A diagonal pattern is visible in \gps\
which is not observed in the PSF magnitudes.  While the per-pixel
scatter is somewhat (10\% to 20\%) higher for these aperture
magnitudes than for the PSF magnitudes
(Table~\ref{table:sigmas.by.filter}), a structure with the amplitude
of the PSF magnitude tree-rings would certainly have been obvious.

\subsection{Astrometric Residuals}

\begin{figure*}[htbp]
\begin{center}
\parbox[b]{\figwidth}{\includegraphics[width=\figwidth]{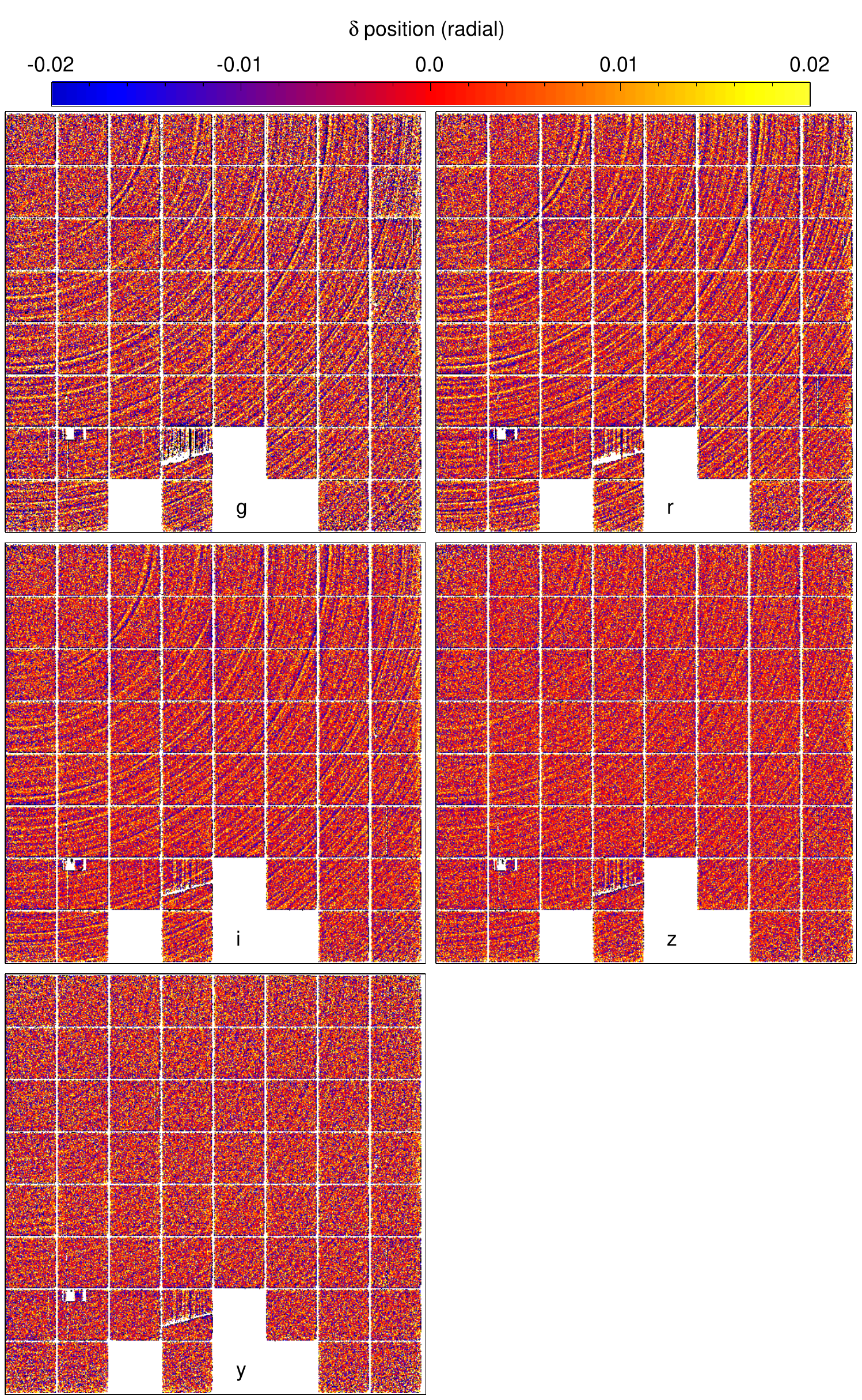}}
\hspace{\jumpleft}
\parbox[b]{\capwidth}{
\caption{Astrometric residuals of the displacement in the radial
  direction, relative to the chip coordinate -5,4960 (upper left
  corner), by filter (\grizy).  White boxes are GPC1 cells which have
  been masked due to poor response.  Superpixels representing regions
  of $10\times10$ pixels are used to determine the median deviation
  for measurements at the given chip pixel location compared with the
  average astrometry for the given
  object. } \label{fig:astrom.by.filter}}
\end{center}
\end{figure*}

Figure~\ref{fig:astrom.by.filter} shows a similar type of measurement
for astrometric residuals.  To generate this plot, we use the same
selection of measurements for astrometry as for photometry.  In this
case, we extract the residual in both the RA and DEC directions
($\delta RA = \overline{RA} - RA_i$, $\delta DEC = \overline{DEC} -
DEC_i$) and rotate these values to the chip coordinate system ($\delta
X,\delta Y$) using our knowledge of the chip orientation on the sky.
We again exclude as bad any measurement with $|\delta X|$ or $|\delta
Y| > 0.5$ arcsec before measuring the median values for each
superpixel.  We have determined the approximate center of the circular
tree-ring pattern as (-5,4960) for this particular chip based on the
pattern of the X astrometry displacements.  Using this coordinate as
the center of the pattern, we have converted the $\delta X,\delta Y$
offsets into $\delta R,\delta \theta$ measurements ($\delta R$ :
radial component away from the center of the pattern, $\delta \theta$
: tangential component).

Figure~\ref{fig:astrom.by.filter} shows the 2D patterns of $\delta R$
for each filter (\grizy).  The dynamic range of the color scale is
from -20 to +20 milliarcseconds for all 5 plots.  A tree-ring
pattern is visible for all five filters, with systematic structures
following a circular pattern centered on the chip corner; the finging
pattern is not apparent in the \yps\ astrometry.  The per-pixel
standard deviations of these plots are listed in
Table~\ref{table:sigmas.by.filter}.  The signal-to-noise of these
structures is again somewhat weak, but the pattern is clearly visible
in these figures.

\subsection{Flat-field Structures}

\begin{figure*}[htbp]
\begin{center}
\parbox[b]{\figwidth}{\includegraphics[width=\figwidth]{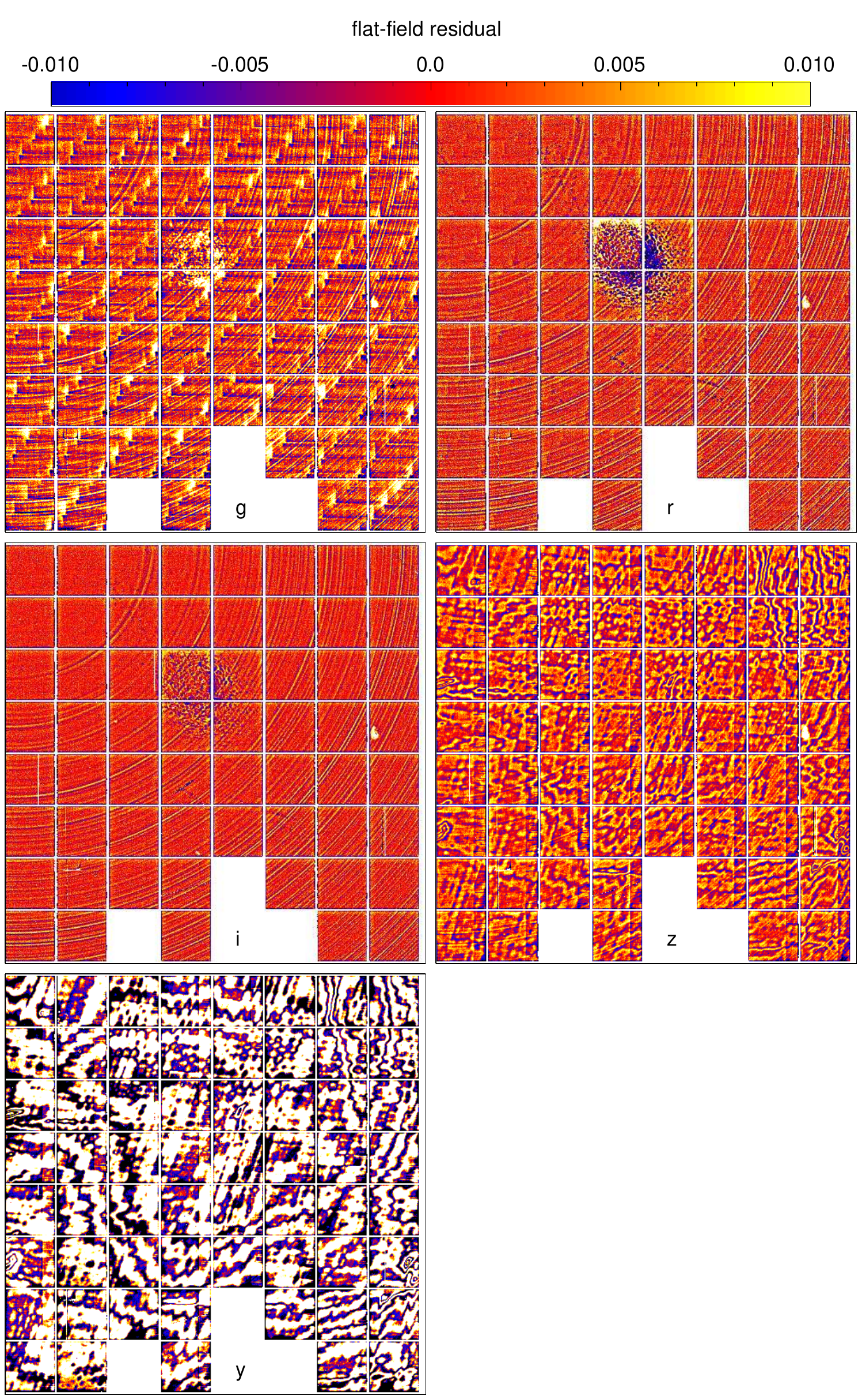}}
\hspace{\jumpleft}
\parbox[b]{\capwidth}{
\caption{Flat-field high-frequency structues, by filter (\grizy).
  White boxes are GPC1 cells which have been masked due to poor
  response.  Flat-field images generated using a tunable laser have
  been combined (see text); a smoothed version has been subtracted to
  high-pass the response.  Flat-field pixels are averaged for
  $10\times10$ superpixels. } \label{fig:flats.by.filter}}
\end{center}
\end{figure*}

Figure~\ref{fig:flats.by.filter} shows the high-spatial-frequency
structures in the flat-field images.  For this measurement, we have
used a set of monochromatic flat-field images obtained with a tunable
laser.  The laser is used to illuminate our flat-field screen which is
then observed by the PS1 telescope.  These flat-field images were
obtained 2011 Feb 09 as part of a campaign to study the PS1 system
response \citep{2012ApJ...750...99T}.  Flats were obtain in a set of
4nm steps sampling the spectral response curve of each filter.  To
enhance the signal-to-noise, we have median-combined a set of 6 flats
at the wavelength center of the corresponding filter.

In order to mask pixels which do not flatten well, we generate a copy
of the image smoothed with a Gaussian kernel with $\sigma = 1.5$
pixels.  Any pixels in the smoothed image which deviate from the
median value in the image by more than 4 standard deviations are
masked.  We generate the superpixel image by averaging the unmasked
pixels associated with each superpixel.  

Figure~\ref{fig:flats.by.filter} shows the superpixel images for the
flat-fields in the five filters.  These flat-field images are
displayed as fractional deviations relative to the median flat-field
image and can thus be compared to the magnitude residuals.  When
flattening an image, these flat-fields would be divided into the flux
of the raw image.  The residuals are thus defined in the sense that a
positive feature in these flats which does {\em not} represent a real
quantum efficiency deviation would induce a {\em reduction} in the
measured flux in those pixels, and thus a {\em negative} deviation in
$\delta m_{psf}$ as defined above.  The dynamic range of the color
scale in these plots is -0.01 to +0.01.  The tree-ring pattern is
strong in the (\gps,\rps,\ips) images, but nearly swamped by fringing
in \zps, and completely lost to finging in \yps.  A diagonal banding
pattern is also seen in \gps: this feature is thought to be due to
the lithography process used to generate the CCD.  A blob can also
been seen covering 4 cells near the center of this chip; this is
apparently a deposit of some kind on the detector.  Both of the latter
two effects behave like quantum efficiency variations and are removed
well by standard flat-field techniques.  Note that a small amount of
the diagonal banding pattern remains in the aperture magnitude
residuals for \gps.  For the rest of this article, we ignore these
features and concentrate on the tree-ring features.

In order to suppress the large-scale structures for a quantitative
analysis of the tree rings, we high-pass filter the superpixel image
by subtracting a copy smoothed with a Gaussian of $\sigma = 3.0$
superpixels.

\subsection{Second Moments}

\begin{figure*}[htbp]
\begin{center}
\parbox[b]{\figwidth}{\includegraphics[width=\figwidth]{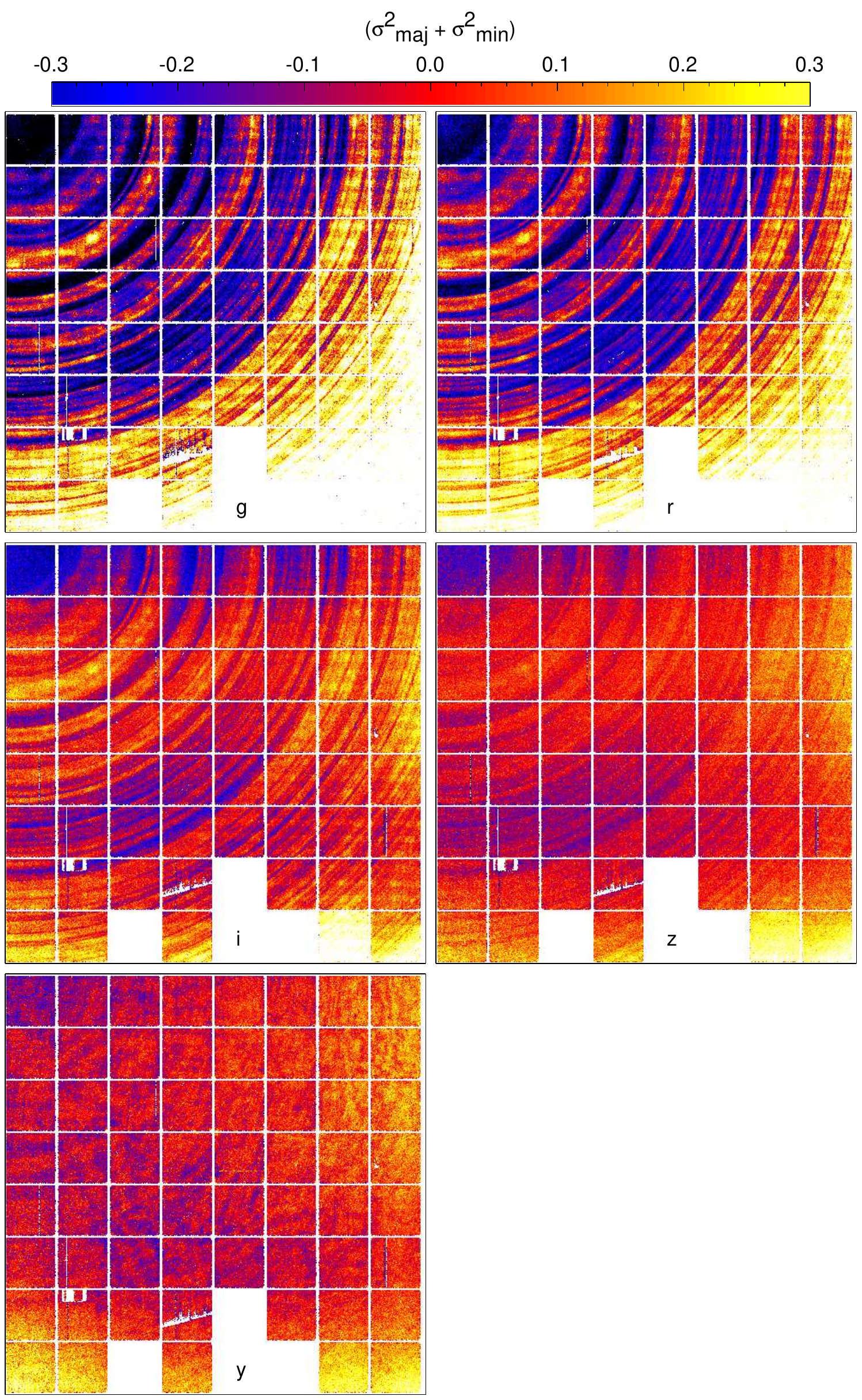}}
\hspace{\jumpleft}
\parbox[b]{\capwidth}{
\caption{Average residual smear variations, by filter (\grizy).  White
  boxes are GPC1 cells which have been masked due to poor response.
  The residual smear ($\sigma^2_{\mbox{major}} + \sigma^2_{\mbox{minor}}$) has been
  determined after the after PSF second moments have been subtracted
  for each image; these values are averaged for each $10\times10$
  superpixels.  } \label{fig:smear.by.filter}}
\end{center}
\end{figure*}

\begin{figure*}[htbp]
\begin{center}
\parbox[b]{\figwidth}{\includegraphics[width=\figwidth]{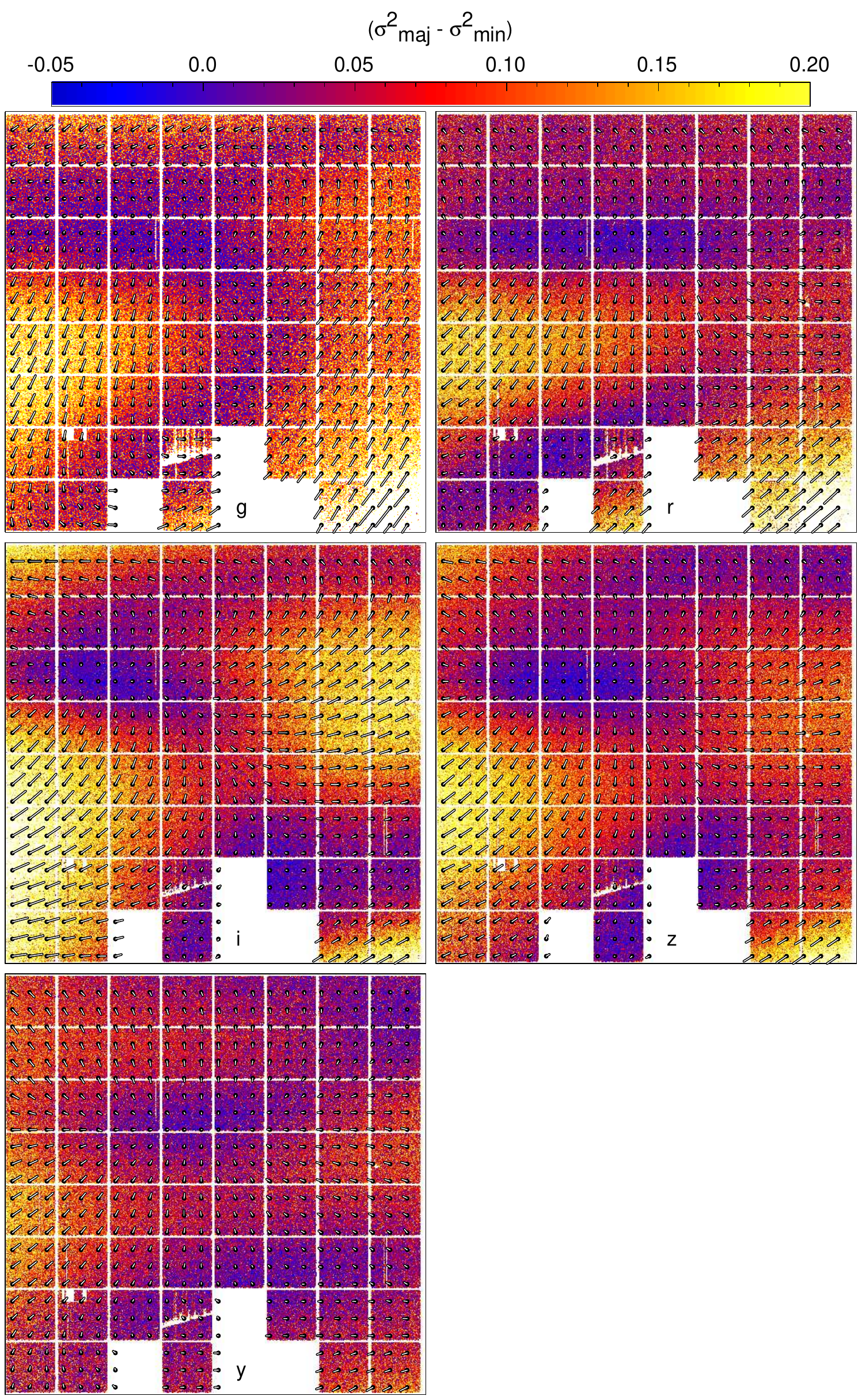}}
\hspace{\jumpleft}
\parbox[b]{\capwidth}{
\caption{Average residual shear variations, by filter (\grizy).  White
  boxes are GPC1 cells which have been masked due to poor response.
  The residual shear ($\sigma^2_{\mbox{major}} - \sigma^2_{\mbox{minor}}$) has been
  determined after the after PSF second moments have been subtracted
  for each image; these values are averaged for each $10\times10$
  superpixels.  } \label{fig:shear.by.filter}}
\end{center}
\end{figure*}

During the image analysis, the second moments are measured for all
stars.  The values can be used to assess changes in the shape of stars
on the image.  To measure changes in the shapes, we have extracted the
second moments for all stellar detections, subject to the same
selections as for the photometry and astrometry residuals (good stars,
multiple detections).  The second moments are measured with a Gaussian
weighting function, with the $\sigma_{w}$ scaled by the PSF size so
that the $\sigma$ measured for PSF stars is \approx 65\% of
$\sigma_{w}$.  (Note that, since the measured $\sigma$ of stellar
objects is biased down by the weighting function, this is not quite
the same as having $\sigma_{w} = 1.6$ times the true PSF $\sigma$; see
discussion in \citealt{magnier2017.analysis}).  For each stellar
detection, we extract the values $M_{xx,xy,yy} = \sum F_i w_i (x^2, x
y, y^2) / \sum F_i w_i$.  For each exposure, we find the median second
moments for PSF objects on this chip (XY40) and subtract those median
values from the instantaneous measurements of $M_{xx,xy,yy}$.  We then
determine the median of the residual second moments for each
superpixel, resulting in 3 images ($\delta M_{xx,xy,yy}$) for each
filter.

Using the second moment images, we can construct certain interesting
combinations, inspired by discussions of lensing measurements \citep{1995ApJ...449..460K}:
\begin{eqnarray}
e_0 & = & \delta M_{xx} + \delta M_{yy}  \\ 
e_1 & = & \delta M_{xx} - \delta M_{yy}  \\
e_2 & = & \sqrt{e_1^2 + 4 \delta M_{xy}}
\end{eqnarray}
For a 2D Gaussian profile with an elliptical contour, these values are
related to the shape of the elliptical contour as follows:
\begin{eqnarray}
e_0 & = & \sigma^2_{\mbox{major}}  + \sigma^2_{\mbox{minor}} \\
e_1 & = & (\sigma^2_{\mbox{major}}  - \sigma^2_{\mbox{minor}}) \cos (2 \theta) \\
e_2 & = & \sigma^2_{\mbox{major}}  - \sigma^2_{\mbox{minor}}
\end{eqnarray}
Where $\sigma_{\mbox{major}}$ and $\sigma_{\mbox{minor}}$ are the
major and minor axis dimensions of the ellipse and $\theta$ is the
position angle.  Thus, $e_0$ is a measurement of the change in the
size of the stellar PSFs as a function of position in the detector
(``smear''), $e_2$ is a measurement of the change in ellipticity of
the stellar PSFs (``shear''), and we can determine the angle of the
PSF ellipticity from the $e_1$ term.

Figure~\ref{fig:smear.by.filter} shows the spatial trend of the smear,
$e_0$.  The dynamic range of these images is -0.3 to +0.3 pixel$^2$. A
tree-ring pattern is visible for all 5 filters, though \yps\ is
dominated by the fringing pattern.  Structures with relatively low
spatial frequencies can also be seen.

Figure~\ref{fig:shear.by.filter} shows the spatial trend of the shear,
$e_2$.  This value is positive definite and is plotted with a color
scale ranging from -0.02 to 0.22 pixel$^2$.  Overlayed on
Figure~\ref{fig:shear.by.filter} is a set of vectors representing the
ellipse orientation as a function of postion.  The length of the
vectors corresponds to the value of $e_2$.  The tree-ring structure is
{\em not} apparent in this figure for any filter.  The spatial
variations are low-frequency and unrelated to the radial trend from
the upper-left corner.

\subsection{Correlations Between Tree-Ring Patterns}

\begin{figure*}[htbp]
\begin{center}
\parbox[b]{\figwidth}{\includegraphics[width=5.0in]{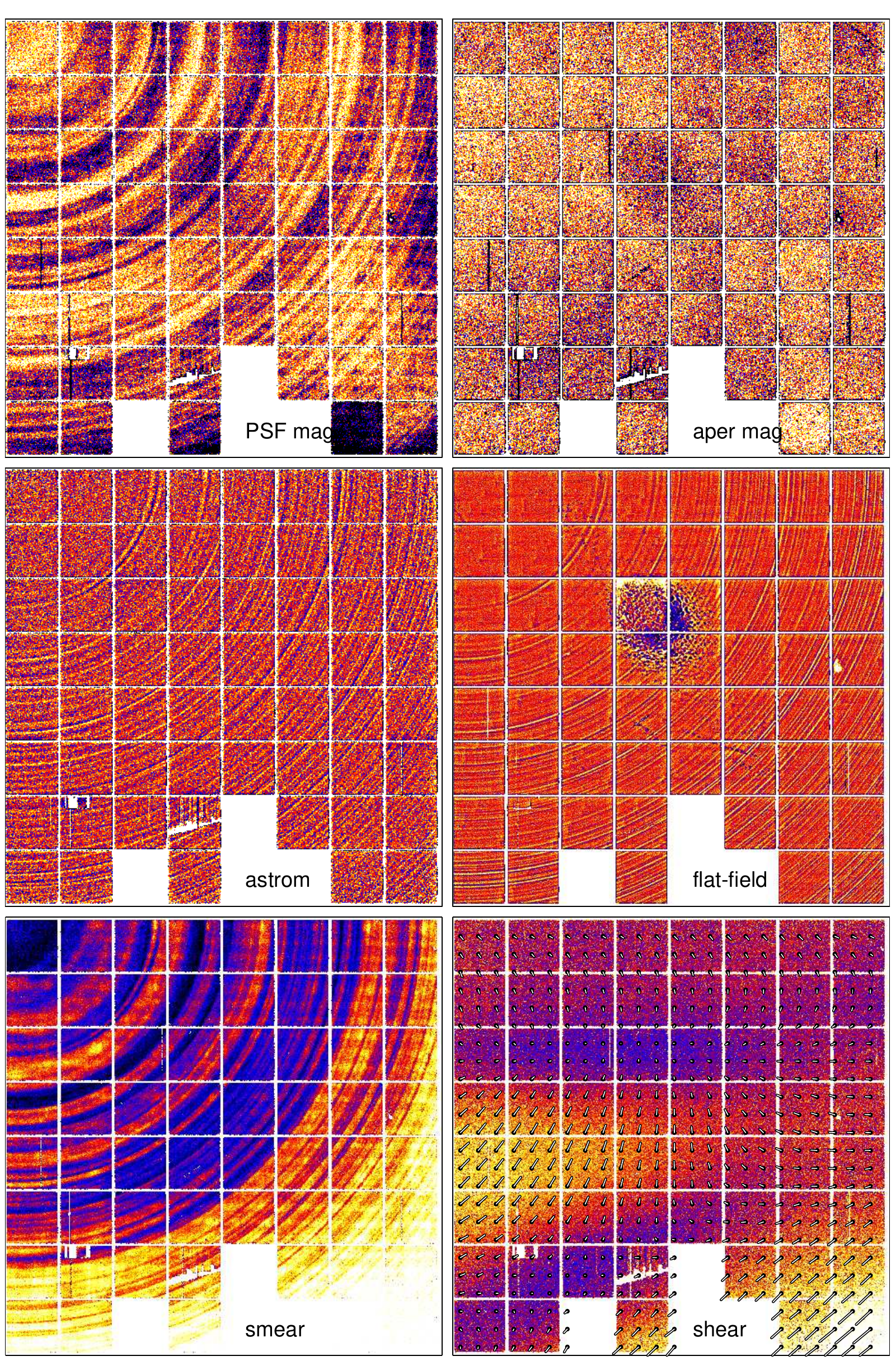}}
\caption{All 6 measured effects for \rps.  This figure illustrates the
  different spatial structure observed for each of the 6 patterns
  measured in this work.  The PSF magnitude (upper-left) and smear
  residuals (lower-left) have a very clear common tree-ring structure,
  while the astrometric residual (middle-left) and flat-field
  residuals (middle-right) have their own common tree-ring pattern with
  much higher frequencies than the previous two effects.  Aperture
  magnitude (upper-right) and shear residuals (lower-right) do not
  show a strong signal consistent with either of the two patterns.} \label{fig:all.effects.rband}
\end{center}
\end{figure*}

\begin{table}
\begin{center}
\caption{Systematic Trends : Correlations by filter\label{table:correlation.by.filter}}
\begin{tabular}{|l|rrrr|}
\hline
{\bf Filter} & {\bf smear} & {\bf psf mags} & {\bf astrom} & {\bf flat} \\
\hline
\gps & 1.00 & 1.00 &  1.00 & 1.00 \\ 
\rps & 0.78 & 0.84 &  0.84 & 0.76 \\
\ips & 0.40 & 0.50 &  0.66 & 0.64 \\
\zps & 0.16 & 0.26 &  0.37 & 0.33 \\
\yps & 0.10 & 0.10 &  0.25 & 0.30 \\
\hline
\end{tabular}
\end{center}
\end{table}

Tree-ring patterns are clearly seen in 4 of the measurement types
above: the PSF photometry, the astrometry, the flat-field, and the
smear terms.  As discussed above, the signal-to-noise per pixel in the
plots of the systematic trends is relatively low (\approx 1.0).  While
the tree-ring patterns are apparent in many of these figures,
there are also some other systematic structures which may degrade the
signal further.

To quantitatively compare the tree-ring trends between filters and
between the types of measurements, we need to measure the tree-ring
structure explicitly and filter out the other effects if possible.  To
do this, we have applied a high-pass filter to all of the relevant
images (PSF photometry residuals, astrometric residuals in the radial
direction, flat-field residuals, and second moment smear terms) to
remove unrelated spatial structures.  We have then measured the median
of the signal in radial bins centered on (-5,4960) across an arc from
$\phi$ = -20\degrees\ to -50\degrees (as measured relative to the top
row of the images).  We have selected a small fraction of the arc to
minimize the error associated with the choice of the pattern center
and to avoid several bad cells near the bottom of the chip.



For a given type of measurement, the systematic effect is strongly
correlated between filters.  The strongest correlation is the smear
term: Figure~\ref{fig:smear.trends} shows the correlation of the smear
pattern between \gps\ and the other four filters. Even \yps\ is
strongly correlated with \gps\ despite the presence of the fringe
pattern.  PSF photometric residuals are also correlated between
filters, as shown in Figure~\ref{fig:psfmag.trends}.  Here, the
\yps\ correlation with \gps\ is quite weak: the fringing pattern
dominates the tree rings for PSF photometry.  The radial component of
the astrometric residual is also well correlated between filters, with
no loss of correlation due to fringing in \yps. Finally, the
flat-field residuals are generally correlated between filters, but
both \zps\ and \yps\ are affected by fringing.  For \yps, the
correlation is completely washed out by the very strong fringing
pattern.

For all four types of measurements, the slope of the fitted lines are
listed in Table~\ref{table:correlation.by.filter}.  There is a
consistency in the trend from \gps, with the strongest systematic
tree-ring effects to \yps, with the weakest effects.  Note that the
second moment smear and astrometry terms have different relative
strength in \yps\ compared with \gps.

\def\figwidth{6.5in}
\begin{figure*}[htbp]
\begin{center}
\includegraphics[width=\figwidth]{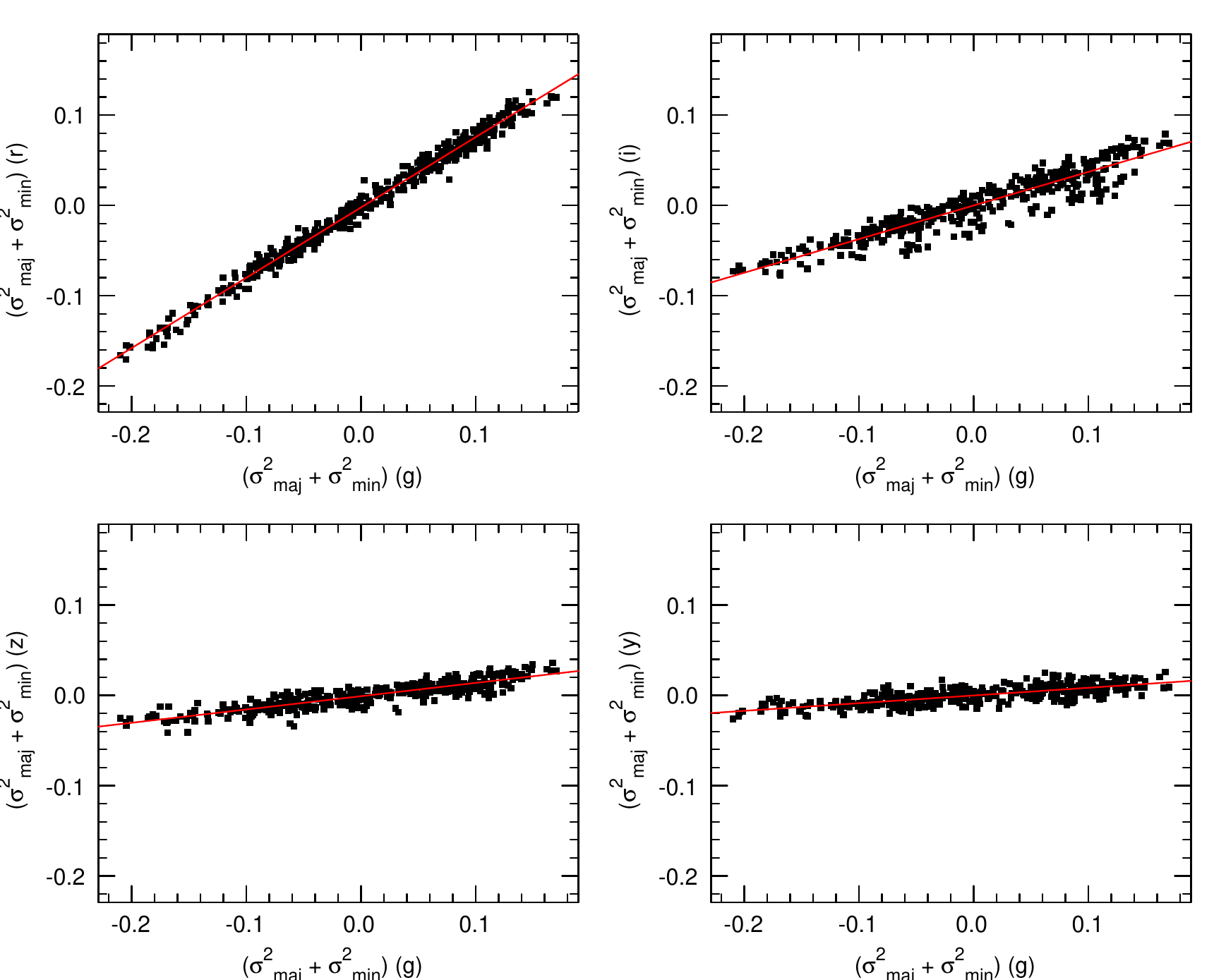}
\caption{Correlation of the smear ($\sigma^2_{\mbox{major}} +
  \sigma^2_{\mbox{minor}}$) signal in \gps\ with the other 4 bands:
  \rps\ (upper-left),  \ips\ (upper-right), \zps\ (lower-left), \yps\ (lower-right).
} \label{fig:smear.trends}
\end{center}
\end{figure*}

\def\figwidth{6.5in}
\begin{figure*}[htbp]
\begin{center}
\includegraphics[width=\figwidth]{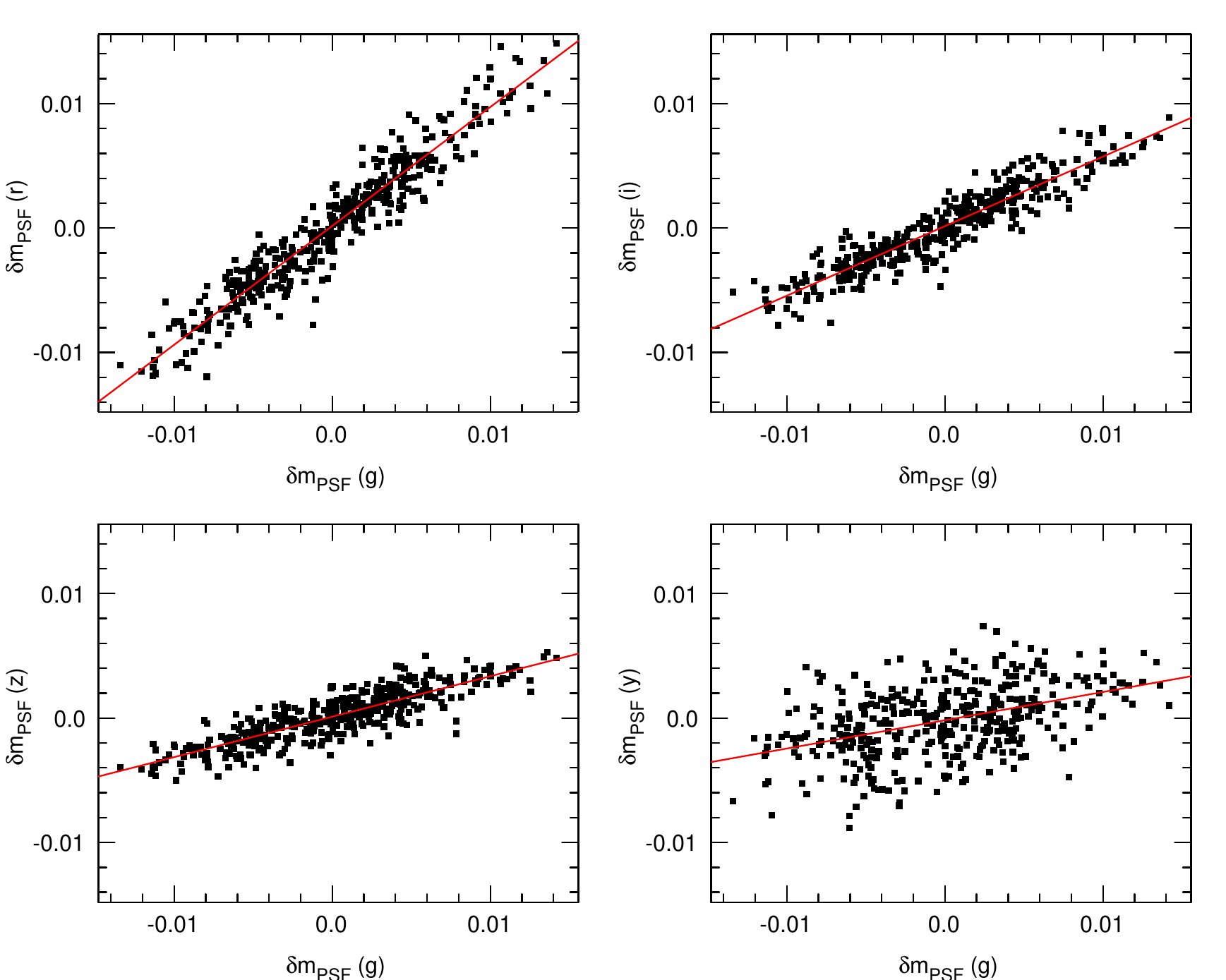}
\caption{Correlation of the PSF magnitude residuals ($\delta m_{psf}$)
  in \gps\ with the other 4 bands: \rps\ (upper-left), \ips\
  (upper-right), \zps\ (lower-left), \yps\ (lower-right).
} \label{fig:psfmag.trends}
\end{center}
\end{figure*}

\def\figwidth{6.5in}
\begin{figure*}[htbp]
\begin{center}
\includegraphics[width=\figwidth]{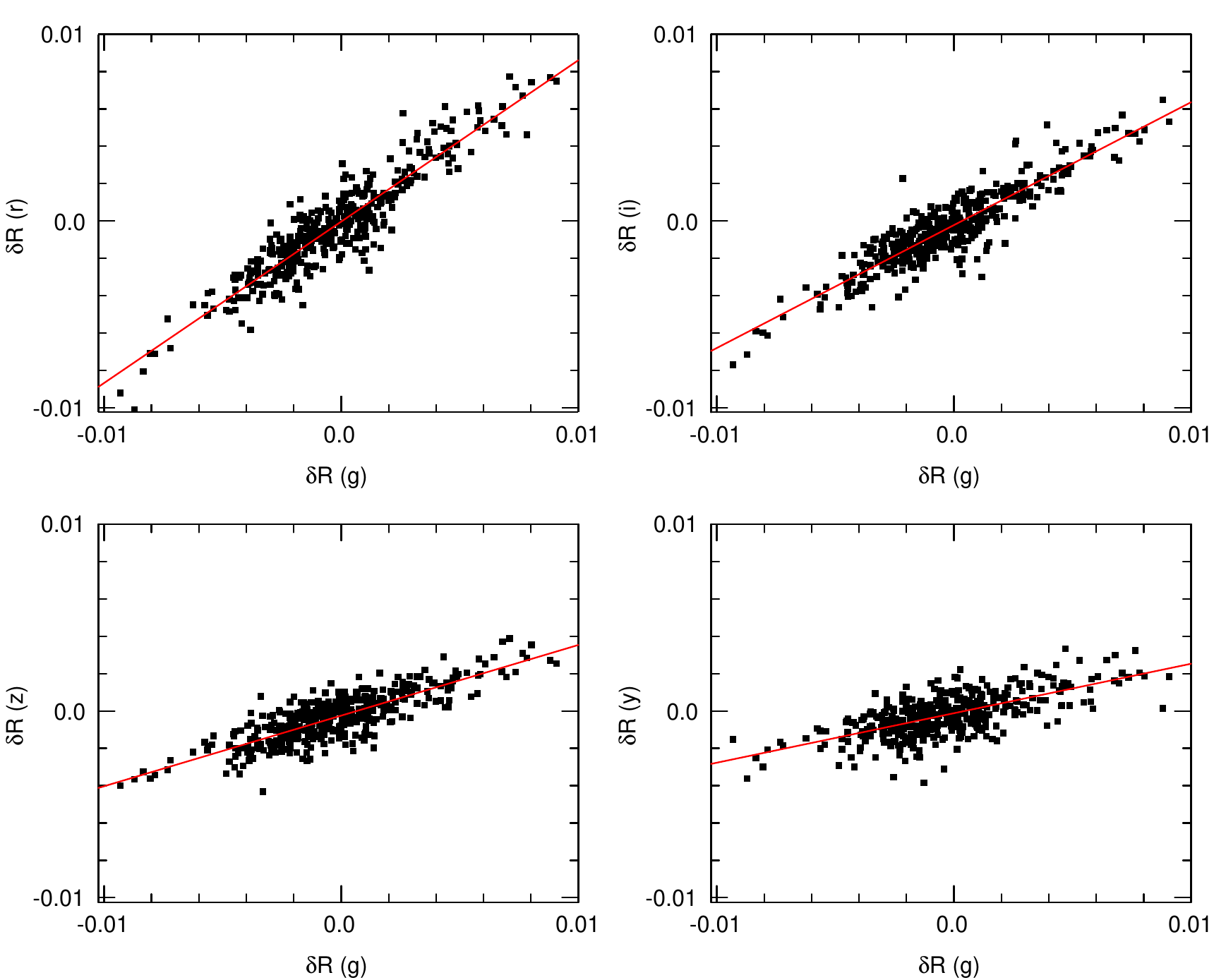}
\caption{Correlation of the radial astrometric residual displacement ($\delta R$)
  in \gps\ with the other 4 bands: \rps\ (upper-left), \ips\
  (upper-right), \zps\ (lower-left), \yps\ (lower-right).
} \label{fig:astrom.trends}
\end{center}
\end{figure*}

\def\figwidth{6.5in}
\begin{figure*}[htbp]
\begin{center}
\includegraphics[width=\figwidth]{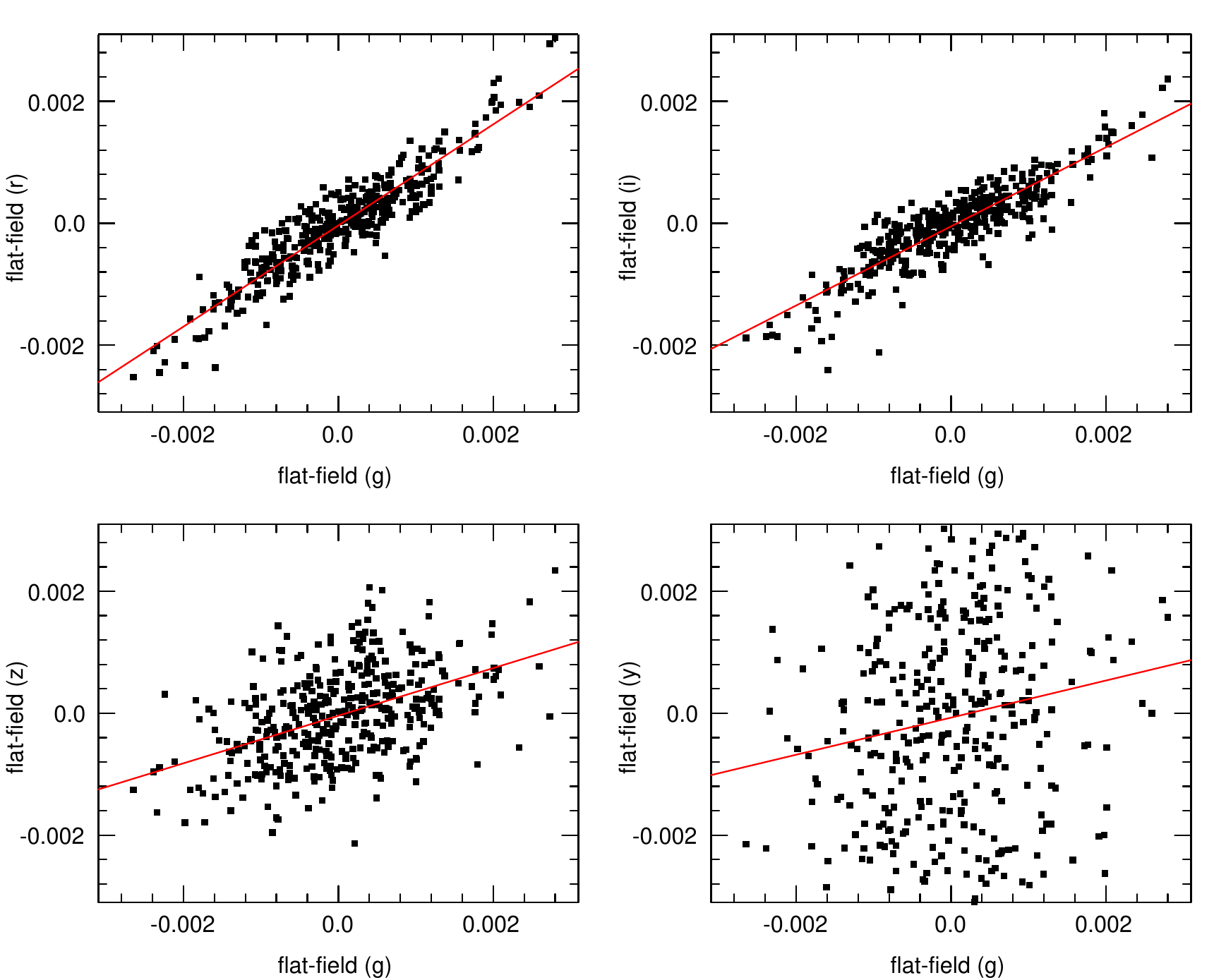}
\caption{Correlation of the flat-field tree-ring structures in \gps\
  with the other 4 bands: \rps\ (upper-left), \ips\ (upper-right), \zps\
  (lower-left), \yps\ (lower-right).  } \label{fig:flat.trends}
\end{center}
\end{figure*}

An important question is the relationship of the tree-ring
pattern between the different types of measurements.  Different models
for the tree-ring structures make different predictions about the
correlations between different effects.  Note the very different
spatial structure between the different measurements in a given
filter: the radial variations do not all follow the same patterns.
Instead, we find the following relationships hold:

First, the PSF magnitude residuals and the second-moment smear trends
are strongly anti-correlated: regions which have larger PSFs than the
mean tend to have smaller measured PSF fluxes than the mean (note that
$\delta m_{psf}$ is defined so that positive values correspond to
larger fluxes).  These trends are shown in
Figure~\ref{fig:smear.vs.psfmag}.  

Second, the radial derivative of the smear is anti-correlated with the
radial component of the astrometric residuals: $\frac{\partial
  (\sigma^2_{major} + \sigma^2_{minor})}{\partial radius} \sim \delta
R$ (see Figure~\ref{fig:dsmear.vs.astrom}).

Finally, the radial derivative of the radial component of the
astrometric residual is anti-correlated with the flat-field residual
errors: $\frac{\partial \delta R}{\partial radius} \sim \delta flat$
(see Figure~\ref{fig:dastrom.vs.flat}).  This last relationship is
somewhat weakly measured.  Because of the periodic nature of the tree
rings, it is also difficult to be completely certain that the
flat-field is proportional to the derivative of the astrometry
residual, rather than the astrometry residual being proportional to
the derivative of the flat-field.  The correlation is somewhat weaker
for derivative of the flat-field vs astrometry residual.  The
correlation is very weak between the flat-field and the astrometry
residual values without a derivative.  We are convinced that we have
the sense of the derivative correct by examination of specific
features in each image.

\begin{table}
\begin{center}
\caption{Systematic Trends : Correlations between trends\label{table:correlation.by.trend}}
\begin{tabular}{|l|rrr|}
\hline
{\bf Filter} & {\bf psf mags} & {\bf $\grad$ smear} & {\bf $\grad$ astrom} \\
             & {\bf vs smear} & {\bf vs astrom}     & {\bf vs flat}        \\
\hline
\gps & -0.056 & -0.060 & -0.47  \\ 
\rps & -0.071 & -0.073 & -0.45  \\
\ips & -0.077 & -0.095 & -0.45  \\
\zps & -0.082 & -0.078 & -0.17  \\
\hline
\end{tabular}
\end{center}
\end{table}

\def\figwidth{6.5in}
\begin{figure*}[htbp]
\begin{center}
\includegraphics[width=\figwidth]{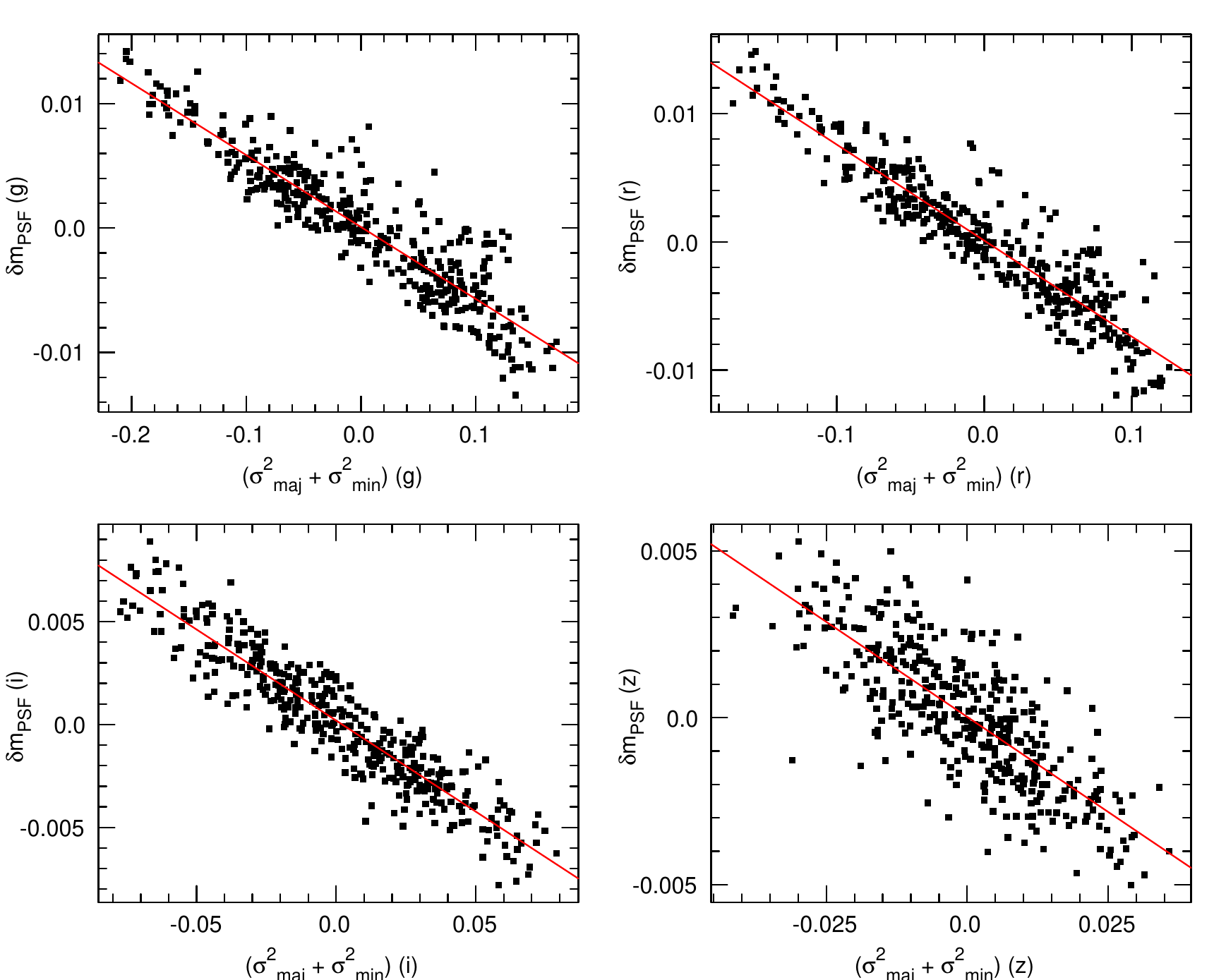}
\caption{Correlation of the PSF magnitude residuals ($\delta m_{PSF}$)
  with the smear ($\sigma^2_{\mbox{major}} + \sigma^2_{\mbox{minor}}$)
  signal for \gps\ (upper-left), \rps\ (upper-right), \ips\ (lower-left),
  \zps\ (lower-right).
} \label{fig:smear.vs.psfmag}
\end{center}
\end{figure*}

\def\figwidth{6.5in}
\begin{figure*}[htbp]
\begin{center}
\includegraphics[width=\figwidth]{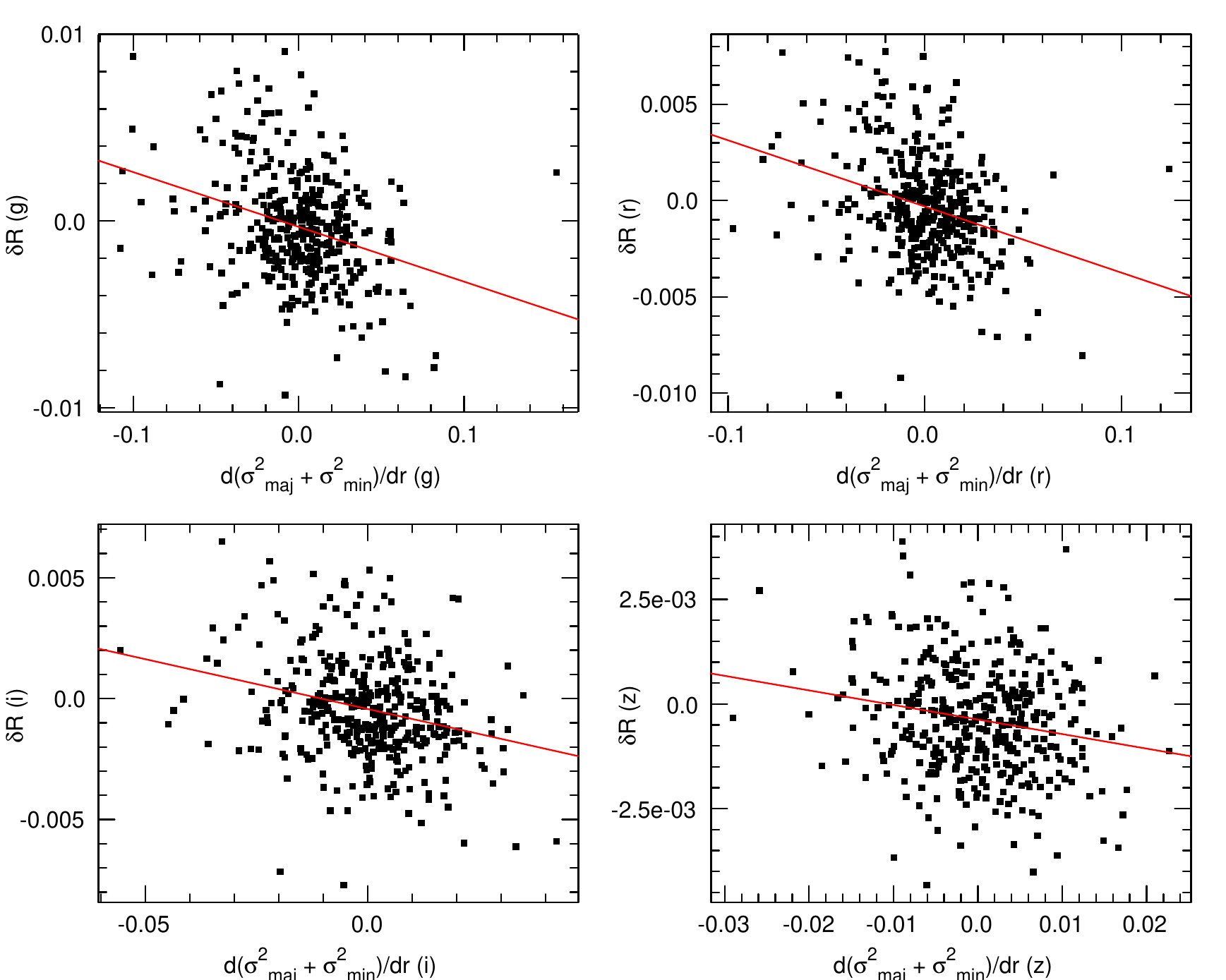}
\caption{
Correlation of the radial astrometric residual displacement ($\delta
R$) with the derivative of the smear ($\partial
\sigma^2_{\mbox{major}} + \sigma^2_{\mbox{minor}}$) signal with
respect to the radial postion for \gps\ (upper-left), \rps\
(upper-right), \ips\ (lower-left), \zps\ (lower-right).
} \label{fig:dsmear.vs.astrom}
\end{center}
\end{figure*}

\def\figwidth{6.5in}
\begin{figure*}[htbp]
\begin{center}
\includegraphics[width=\figwidth]{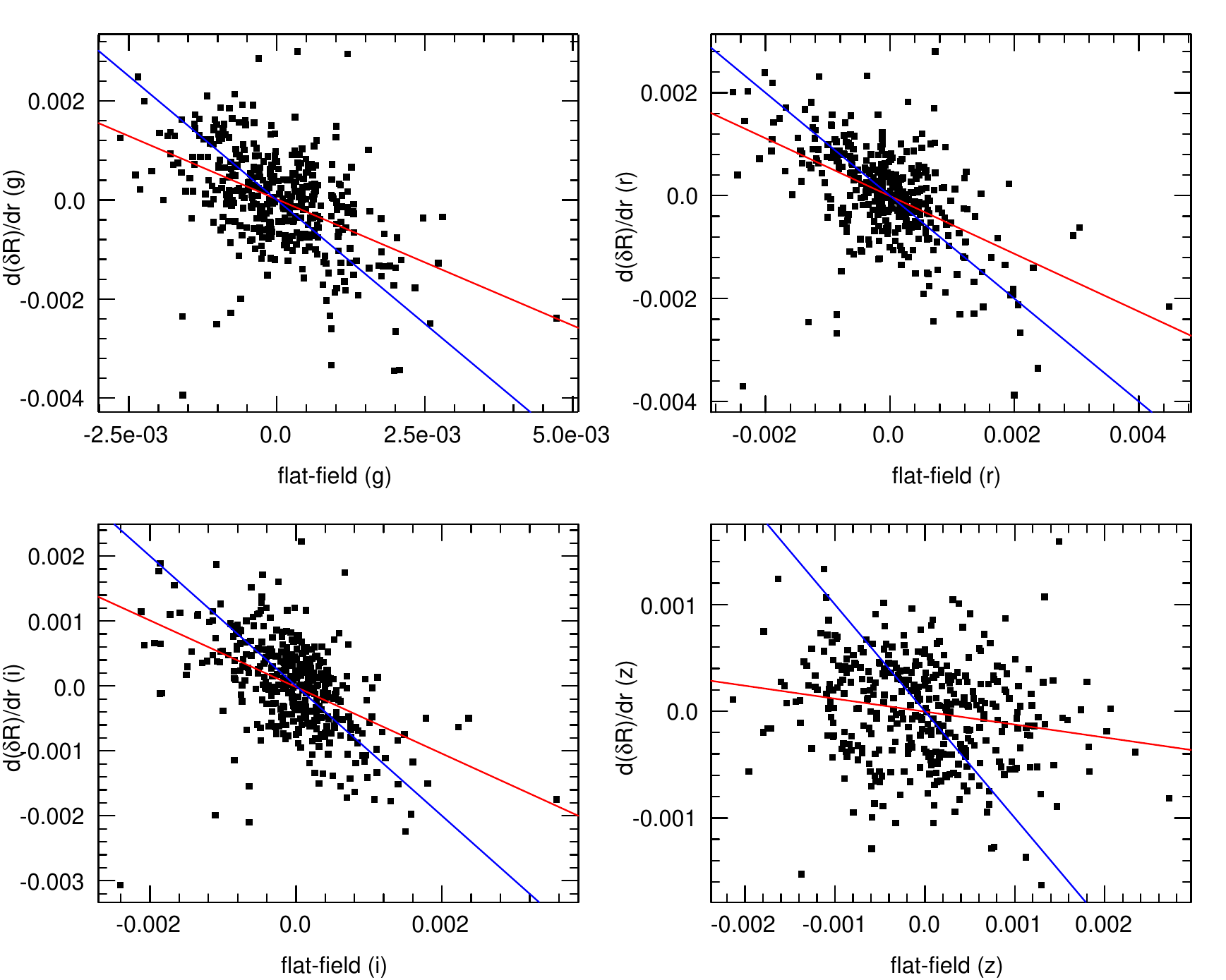}
\caption{
Correlation of the derivative of the radial astrometric residual
displacement ($\delta R$) with respect to the radial position with the
flat-field tree-ring signal for \gps\ (upper-left), \rps\ (upper-right),
\ips\ (lower-left), \zps\ (lower-right).
} \label{fig:dastrom.vs.flat}
\end{center}
\end{figure*}

\section{Discussion}
\label{sec:discussion}

These trends measured above (Section~\ref{sec:tree.rings}) help to
illuminate the underlying causes of these different effects.

First, if we consider the smear pattern
(Figure~\ref{fig:smear.by.filter}), the measurement shows that the
intrinsic sizes of the stellar images are varying in a radial sense
between the different tree-ring regions.  Although images experience
an average image quality (due to seeing and focus) across the chip
which may vary substantially from exposure to exposure, stars landing
in the different tree-ring regions are consistently somewhat
larger or somewhat smaller than that average.

Next, we can explain the correlation between the PSF photometry
residuals and the observed smear (Figure~\ref{fig:smear.vs.psfmag}).
In the photometry analysis, we model the PSF allowing for some spatial
variation in the shape.  However, we have a limited number of stars to
measure any spatial variation.  Thus the 2D variations are sampled on
a very coarse (e.g., $3 \times 3$) grid for each chip: the PSF
parameters may vary smoothly across the chip following the bilinear
interpolation between the $3 \times 3$ grid points.  Thus, the spatial
scale on which we model PSF variations is much larger than the spatial
scale on which PSF variations are actually occuring, as illustrated
by the changes in the smear plot (Figure~\ref{fig:smear.by.filter}).
When the true PSF is larger than the model PSF, our model fits
systematically underestimate the amount of flux in a given object.
Conversely, when the true PSF is smaller, we overestimate the flux -- this
type of offset is a typical effect when mis-estimating the PSF size.
The slope of the trend depends on the mean typical seeing for the
given filter.  For example, the \gps\ seeing is typically 1.3\arcsec,
corresponding to a Gaussian $\sigma$ of 2.15 pixels.  A smearing of
$\sigma^2_{major} + \sigma^2_{minor} = 0.1$ pixels$^2$ would increase
the size by about 0.02 pixels, or 1\%, roughly consistent with the
observed photometric deviation of about 5 to 10 millimags for this
amount of smearing.

The correlation between the flat-field structures and the radial
derivative of the astrometric residual displacements in the radial
direction (Figure~\ref{fig:dastrom.vs.flat}) is consistent with radial
variations in the plate-scale.  The tree-rings observed by DES are
completely attributed to effective plate scale changes.  Effective
plate scale changes result in flat-field deviations because the
flat-field illumination is a source of constant surface brightness.
Pixels see a varying amount of flux depending on their effective area.
This changing plate scale also affects the astrometry since these
variations occur on spatial scales much smaller than the astrometric
model.  In this description of the tree rings, the flat-field
deviations are $-1 \times \frac{\partial \delta R}{\partial r}$.  The
best-fit slopes of our correlations are \approx 0.5, but the
signal-to-noise is rather low.  A slope of -1 appears to be consistent
with our measurements.

The fact that the PSF ellipticity changes are {\em not} correlated
with the tree-ring structure (Figure~\ref{fig:shear.by.filter}) tells us
that, unlike the case for DES, the effective plate-scale changes seen
in the flat-field and astrometry signals are not the dominant cause of
the PSF photometry errors.  Also, the fact that we do not measure
significant aperture photometry errors correlated with the tree rings
confirms this point.  The amplitude of the flat-field errors are 1-2
millimagnitudes, much smaller than the PSF photometry errors, and far
below the pixel-to-pixel noise in the aperture magnitude residuals.
It is likely in our opinion that the plate-scale changes causing the
flat-field and astrometry effects are affecting both the ellipticity
and the aperture magnitudes, but the level of the effect is too small
to see given the other systematic structures (in the shear plot) and
the noise level (in the aperture magnitudes).

Finally, the correlation between the smear structures and the
astrometry residuals shows that these two effects are connected.
Although the correlation is weak in Figure~\ref{fig:dsmear.vs.astrom},
careful inspection of the location of these two tree ring patterns
shows that the locations of the rings in the radial astrometric
residual images occurs at the boundaries between regions with
substantially different values of the smear signal.

We suggest that the underlying connection between all of these
tree-ring effects is the pattern of the doping variations in the
silicon.  As discussed by \cite{2014PASP..126..750P}, the tree-ring
patterns seen by the DES team are caused by lateral electic fields in
the detector silicon (in the plane of the CCD wafer) generated by
variations in the space charges embedded in the silicon, in turn
coming from low-level changes in the doping as the silicon boule is
grown.  We conclude that the astrometric and flat-field variations
seen in our detectors are caused by these same types of doping
variations.  The changes in the smear (and thus the PSF magnitudes)
are apparently also related to the doping variations.  The lateral
electric fields which introduce the astrometry and flat-field
variations occur at the boundary between regions with higher and lower
space charges from the dopant.  Regions with high (or low) space
charge density thus correspond to regions with relatively high (or
low) amounts of smear; the astrometric deviations follow the gradient
between these regions.

We interpret the changes in the smear term as changes in the amount of
charge diffusion as the photoelectrons travel to the bottom of the
pixel well.  The blue filters exhibit the strongest changes in the
amount of smear.  These are also the filters for which the detected
electrons have travelled the longest distance in the silicon, and are
thus most affected by diffusion effects.  Charge diffusion (as opposed
to the charge drift caused by the lateral electric fields) results in
a Gaussian smearing of the stellar profile: as the photoelectrons
migrate from the site where they were generated by the incoming photon
to the bottom of the pixel well, they follow a random walk in the
plane of the detector.  The longer the electrons take to make the
journey down to the bottom of the pixel, the further they are able to
wander from their creation coordinate in the detector.  Following the
discussion in \cite{Holland.2003}, the amount of charge diffusion is
thus related to the velocity of the electrons in the direction of the
optical axis: $\sigma \sim \sqrt{2Dt}$ where $\sigma$ is the size of
the smearing kernel, $t$ is the time required for the electrons to
traverse the thickness of the silicon wafer, and $D$ is the diffusion
coefficient.  The velocity of the photoelectron, and thus the time to
traverse the silicon, is related to the vertical electric fields in
the silicon, which are caused by a combination of the applied voltages
and the distribution of the space charges from the dopant.  As shown
by \cite{Holland.2003}, the charge diffusion is related to the space
charge density by $\sigma \sim \rho^{-\frac{1}{2}}$ (their equation
6).  Regions with high space charge densities increase the migration
speed of the photoelectrons and reduce the amount of charge diffusion
smearing; and vice versa for regions of low space-charge densities.

In summary, the variations in the space-charge density caused by
variations in the dopant result in regions of higher and lower charge
diffusion, and in turn regions with PSF photometry systematic
residuals.  The lateral gradients in the space-charge density induce
lateral electric fields which in turn cause lateral motions of the
photoelectrons, resulting in astrometric and flat-field deviations.

The DES team did not detect these charge diffusion variations.  In
that case, the amplitude of the photometric effects due to the lateral
field are dominant; these include both the modification of the
flat-field as well as PSF fitting errors due to the changing PSF sizes
introduced by the varying effective pixels sizes.  If the smearing
effect reported here were as large for DES compared with the lateral
PSF size changes as they are for GPC1, then the reported PSF
photometry residuals for would have had very different
characteristics.  We conclude that, for DES, the lateral effects are
much larger than the diffusion variations, compared with GPC1.  The
relative amplitude of these two effects depends on the details of the
applied voltages, the amplitude of the space-charge density variations
compared with the typical space-charge density, and the detector
thicknesses.  It is beyond the scope of this article to model these
effects in detail.


\section{Conclusion}

The tree rings observed in the Pan-STARRS GPC1 data show two different
effects, though they are related.  First, the images are experiencing
circularly-symmetric changes in the PSF size correlated with the
tree-ring pattern.  These PSF size changes drive errors in the PSF
photometry on the scale of a few millimagnitudes, and are also
correlated with the tree-ring pattern.  These PSF size changes are
consistent with changes in the charge diffusion, which also introduces
a circularly symmetric smearing.

In addition, there are radial plate-scale changes correlated with the
tree rings.  These plate-scale changes introduce flat-field errors on
the scale of \approx 1 millimagnitude and astrometric errors on the
scale of 2-3 milliarcseconds.  The observed relationship between the
flat-field deviations and the radial derivative of the astrometric
deviations confirms this interpretation \citep[see also discussion
  in][]{2014PASP..126..750P}.

The spatial correlation of the gradient in the smear variations and
the astrometric variations imply that both of these two types of tree
ring effects are related, even though they manifest through different
mechanisms.  We conclude that the variations in both the vertical charge
diffusion and the lateral charge migration are driven by changes
in the electric field structures in the silicon due to the same
variations in the doping structures in the silicon.


The small-scale variations in the charge diffusion observed in the
Pan-STARRS detectors represents a new type of systematic effect in
deep depletion devices.  This feature, if present in other detectors,
could manifest in systematic errors in several ways.  Like in the
Pan-STARRS analysis example, the charge diffusion variations result in
fine-structure in the observed stellar point-spread functions.  For
very precise photometry or morphological analysis, it will be
necessary for the PSF models to account for the extra charge
diffusion.  Unlike the non-uniform pixel-size effects, correction of
the PSF photometry cannot simply be performed as an average flat-field
correction on the measurements after they have been processed.  
The additional smearing acts as a convolution with a Gaussian kernel
of fixed size for a given filter.  The photometry bias is a function
of the fractional change of the PSF size.  Thus, the introduced error
depends on the average PSF for the image in question: an image with
good image quality will suffer larger PSF model errors than an image
with poor image quality.  To account for this effect in a rigorous
way, the analysis should use the measured diffusion variations to
modify the model PSFs as a function of position before they are used
for the image analysis.

The charge diffusion variations may also have an impact on
spectroscopic measurements.  Modern, precise spectroscopic
measurements rely on precise measurements of the stellar line
profiles.  If such an analysis ignores variations in the charge
diffusion, the measured line widths may be systematically biased.

This analysis points to the importance of careful instrumental
characterization, especially for those instruments which are used for
large-scale surveys with largely automatic data analysis systems and
stringent precision goals.

\acknowledgments

The Pan-STARRS1 Surveys (PS1) have been made possible through
contributions of the Institute for Astronomy, the University of
Hawaii, the Pan-STARRS Project Office, the Max-Planck Society and its
participating institutes, the Max Planck Institute for Astronomy,
Heidelberg and the Max Planck Institute for Extraterrestrial Physics,
Garching, The Johns Hopkins University, Durham University, the
University of Edinburgh, Queen's University Belfast, the
Harvard-Smithsonian Center for Astrophysics, the Las Cumbres
Observatory Global Telescope Network Incorporated, the National
Central University of Taiwan, the Space Telescope Science Institute,
the National Aeronautics and Space Administration under Grant
No. NNX08AR22G issued through the Planetary Science Division of the
NASA Science Mission Directorate, the National Science Foundation
under Grant No. AST-1238877, the University of Maryland, and Eotvos
Lorand University (ELTE) and the Los Alamos National Laboratory.


\bibliographystyle{apj}

\end{document}